\documentclass[final,1p,times]{elsarticle} 
\usepackage{graphicx} 
\usepackage{amssymb} 
\usepackage{amsthm} 
\usepackage{lineno} 

\journal{Nuclear Physics A} 
\begin{document} 

\begin{frontmatter} 


\title{$\eta/s$ of a Relativistic Hadron Gas at RHIC: Approaching the AdS/CFT bound?}

\author{Nasser Demir and Steffen A. Bass}

\address{Department of Physics, Duke University, Durham, NC 27708, USA}

\begin{abstract} 
Ultrarelativistic heavy-ion collisions at the Relativistic Heavy Ion Collider (RHIC) are thought to have produced a state of matter called the quark-gluon plasma, characterized by a very small shear-viscosity to entropy-density ratio $\eta/s$, near the lower bound predicted for that quantity by AdS/CFT methods. As the produced matter expands and cools, it evolves through a phase described by a hadron gas with rapidly increasing $\eta/s$. We calculate $\eta/s$ as a function of temperature in this phase both in and out of chemical equilibrium and find that its value poses a challenge for viscous relativistic hydrodynamics, which requires small values of $\eta/s$ in order to successfully describe the collective flow observables at RHIC. We therefore conclude that the origin of the low viscosity matter at the RHIC must be in the partonic phase of the reaction.
\end{abstract} 

\end{frontmatter} 


\section{Introduction}\label{}

Ultrarelativistic heavy ion collisions at the Relativistic Heavy Ion Collider (RHIC) are thought to have produced a Quark Gluon Plasma (QGP) with the characteristics of a near ideal fluid, with large elliptic flow and low shear viscosity to entropy density ratio $\eta/s$ \cite{ideal}.  Relativistic viscous hydrodynamic calculations require low values of $\eta/s$ in order to reproduce the RHIC elliptic flow ($v_2$) data \cite{vhydro1,vhydro2}.  However, current calculations assume a temperature-independent value of $\eta/s$ throughout the entire evolution of the system. The exact value of $\eta/s$ in these calculations is estimated to be in the range of $\eta/s \approx(1-3)/4 \pi$, due to systematic uncertainties, and types of initial conditions used \cite{vreport}.  However, it should be strongly noted that the shear viscosity of matter in a relativistic heavy ion collision is a \textit{time-dependent} quantity.  While the partonic phase of such a collision is expected to have a very low value of $\eta/s$, after hadronization occurs $\eta/s$ is expected to rapidly increase.  Hence, in order to quantitatively constrain the values of $\eta/s$ in the deconfined phase of a relativistic heavy ion reaction, it is necessary to perform a separate calculation of $\eta/s$ in the hadronic phase of the reaction, with a systematic analysis of effects of the system evolving out of chemical equilibrium in the hadronic phase \cite{freezeout}.  In these Proceedings, we calculate $\eta/s$ as a function of temperature in the hadronic phase using the Ultrarelativistic Quantum Molecular Dynamics (UrQMD) model, both in and out of chemical equilibrium \cite{ndprl}.  This suggests a potentially strong time dependence of $\eta/s$ in the hadronic phase of the reaction, and hence that viscous hydrodynamics practitioners would need to incorporate a temperature dependent $\eta/s$ in their simulations.

\section{Model for Hadronic Medium and Viscosity Calculation}\label{}

The model for our hadronic medium is UrQMD, a microscopic transport model based upon the Boltzmann equation.  It contains 55 baryonic species (and their antiparticles) and 32 mesonic species, including strange resonances with the most massive baryonic resonance being the $N^*_{\rm 2250}$ and the most massive mesonic resonance being the $\phi_{\rm 1900}$ \cite{urqmd}.  The cross sections used to calculate the collision integral in solving the Boltzmann equation are parametrized from fits to experimental data for processes that are well known and effective symmetries are used to indirectly infer cross sections for processes not known well experimentally.  The system is initialized with particles uniformly distributed in real space, and it evolves through a sequence of binary elastic and inelastic processes, and the system is forced into equilibrium by confining the particles in our system to a box with periodic boundary conditions.  In order to be able to achieve the proper chemical equilibrium, we disable multiparticle processes, since it is not possible to implement multiparticle collisions using the geometric interpretation of the cross section in UrQMD.  Chemical equilibrium is then verified by comparing hadronic yields to an independent statistical model for a hadron resonance gas (SHARE), and kinetic equilibrium is verified by fitting momentum distributions to a Boltzmann distribution to extract the temperature.

In order to extract the shear viscosity of our system, we employ the Kubo formalism.  The Kubo formalism relates linear transport coefficients to near-equilibrium correlations of dissipative fluxes and treats dissipative fluxes as perturbations to local thermal equilibrium .  The Green-Kubo formula for shear viscosity is 
\begin{equation}
\eta= \frac{1}{T} \int d^3r\int_0^{\infty}\,dt \langle \pi^{xy}(\vec{0},0)\pi^{xy}(\vec{r},t)\rangle_{\rm equil} ,
\end{equation}
where $T$ is the temperature of the system,  $t$ is the post-equilibriation time (the above formula defines \begin{math}t=0\end{math} as the time the system equilibriates), and $\pi^{xy}$ is the shear component of the energy momentum tensor $\pi^{\mu \nu}$ \cite{kubo}.  The expression for the energy momentum tensor $\pi^{\mu \nu}$ is 
\begin{equation}
\pi^{\mu \nu}= \int d^3p \frac{p^{\mu}p^{\nu}}{p^0} f(x,p),
\end{equation}where $f(x,p)$ is the phase space density of the particles in the system.  A representative sample of the shear viscous correlation function is shown on the left panel of Fig. 1.  In order to compuate an integral, we assume an exponential ansatz for the correlator, 
\begin{equation}
\langle \pi^{xy}(0)\pi^{xy}(t)\rangle \propto exp\left(-\frac{t}{\tau_{\pi}}\right),
\end{equation}
so that the viscosity becomes
\begin{equation}
\eta=\frac{V}{T}\langle \pi^{xy}(0)^2\rangle \tau_{\pi}.
\end{equation}

\begin{figure}
\includegraphics[width=0.46\textwidth]{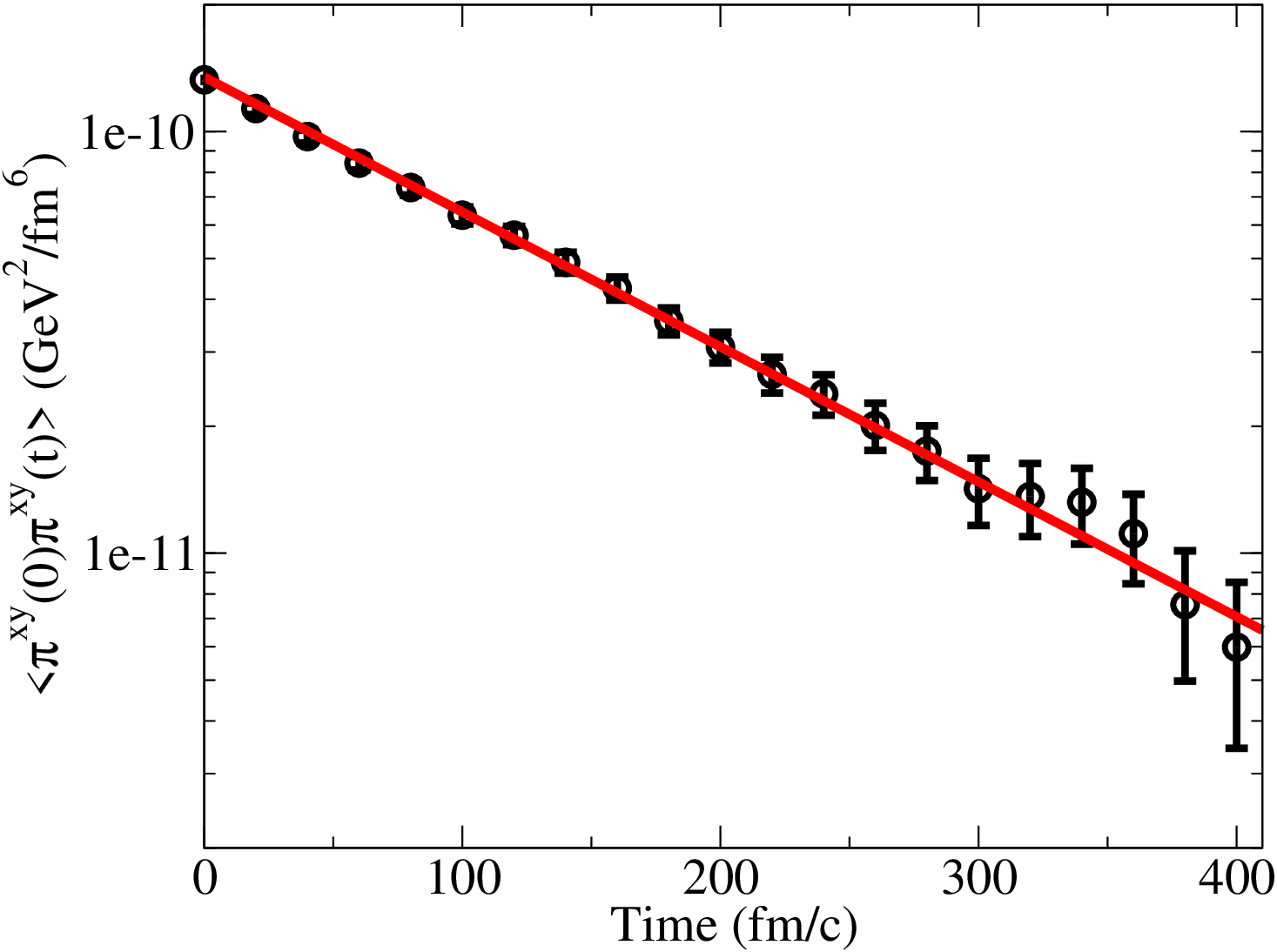}
\includegraphics[width=0.49\textwidth]{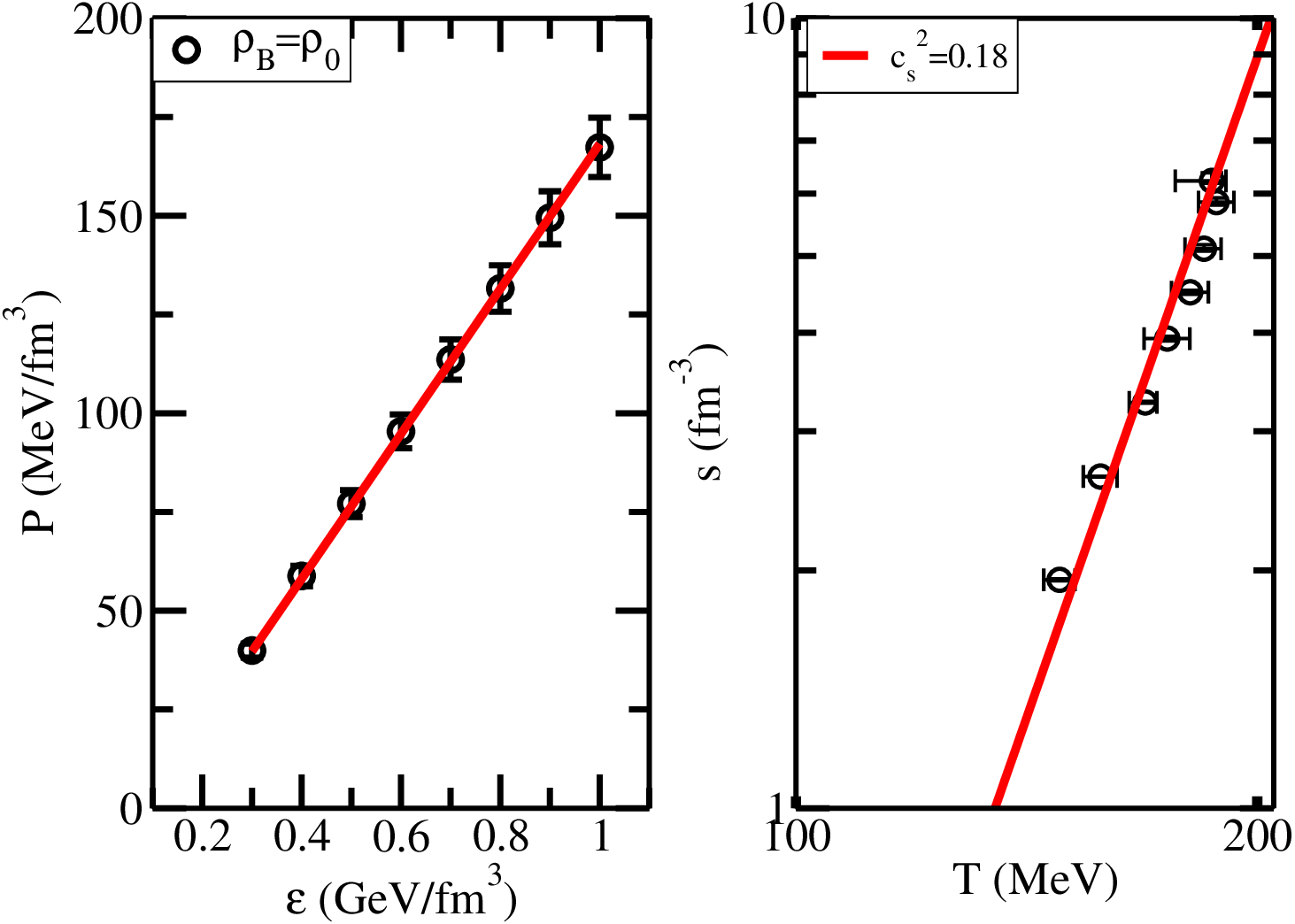}
\caption{LEFT: the shear viscous correlator as a function of time for T=67.9 MeV.  RIGHT: pressure vs. energy density and entropy density versus temperature.}
\end{figure}

\section{Entropy Considerations}

Caution must be taken when extracting the entropy accurately from a microscopic transport model; the system cannot be treated as an ideal gas of massless particles, and specific entropy contributions from particles of different masses are not the same.  We use the Gibbs formula $s=\left(\frac{\epsilon+P-\mu_B\rho_B}{T} \right)$, and extract the energy density, pressure, baryo-chemical potential, and number densities of the relevant chemical species once the system has equilibriated.  We verify the calculation of $s$ via a scaling relation between entropy density and temperature of an equilibriated system at fixed volume: $s \sim T^{\frac{1}{c_s^2}}$, with $c_s$ the speed of sound.  The right panel of Fig. 1 shows pressure as a function of energy density, and a value of $c_s^2=0.18$ extracted from the slope can be used to fit the entropy density versus temperature results in that figure.  This assures us that our calculation is accurate and representative of a hadronic medium of many different hadronic species.    

\section{Results in Full Chemical and Kinetic Equilibrium}\label{}

Using the aforementioned techniques, we present $\eta/s$ as a function of temperature in full equilibrium for zero baryochemical potential in the left panel of Fig. 2. Also illustrated in that figure is the calculation of $\eta/s$ for chiral pions \cite{prakash} and 3 flavor perturbative QCD \cite{amy}. The minimum value found for $\eta/s$ for the equilibrium and zero baryochemical potential case is $\approx$ (0.9-1.0), significantly higher than the KSS bound of $\eta/s \approx$ 0.08. If the minimum value of $\eta/s$ for hadronic matter in the range of hadronic freeze-out indeed occurs at that value, a viscous hydrodynamics fit with a constant value of $\eta/s$ to RHIC data may not yield reliable results for $\eta/s$ in the deconfined phase, since a value of $\eta/s$ of at most 0.24 is needed to reproduce RHIC elliptic flow data \cite{vhydro1}.
\begin{figure}[h]
\centering
\includegraphics[width=0.48\textwidth]{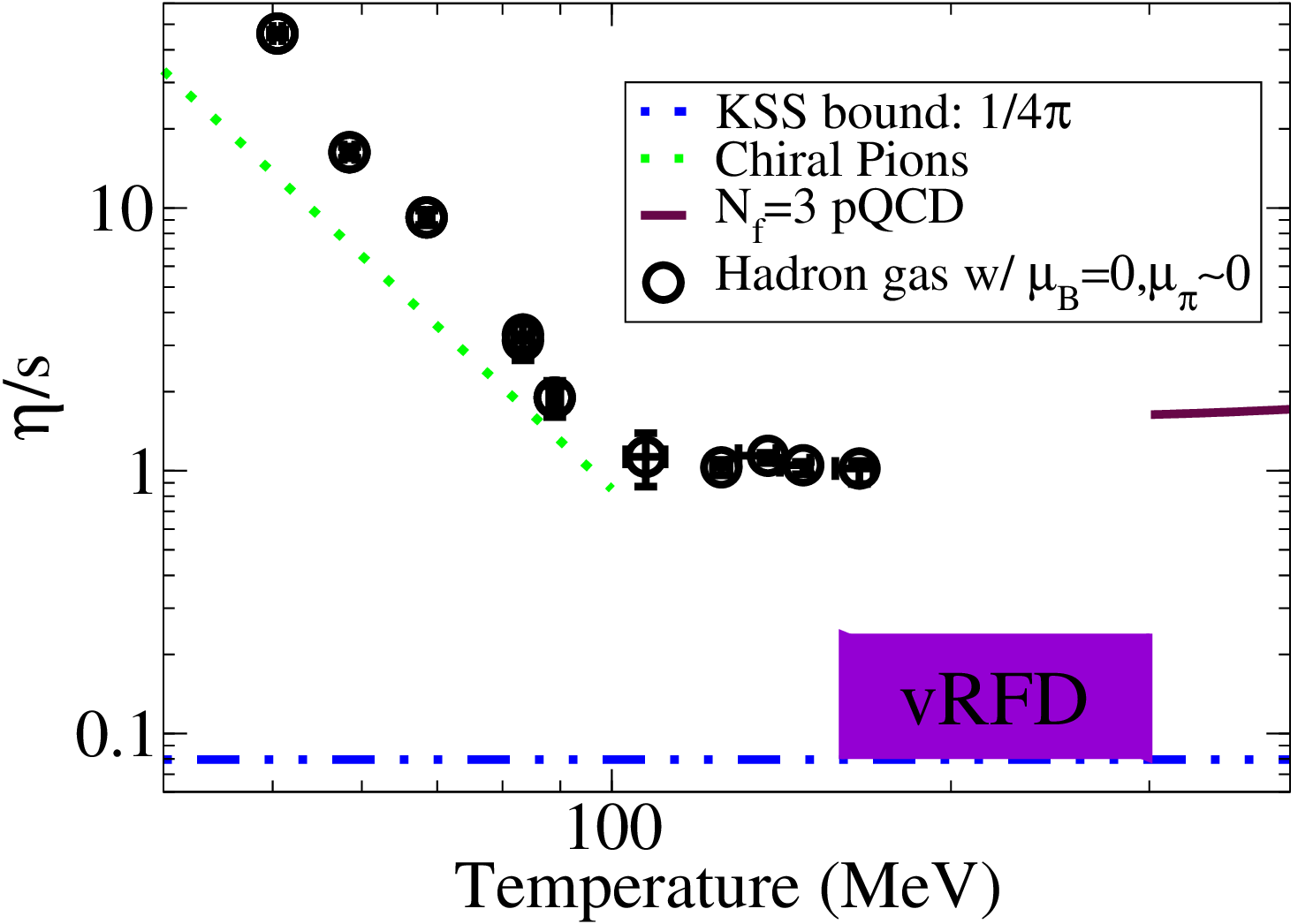}
\includegraphics[width=0.48\textwidth]{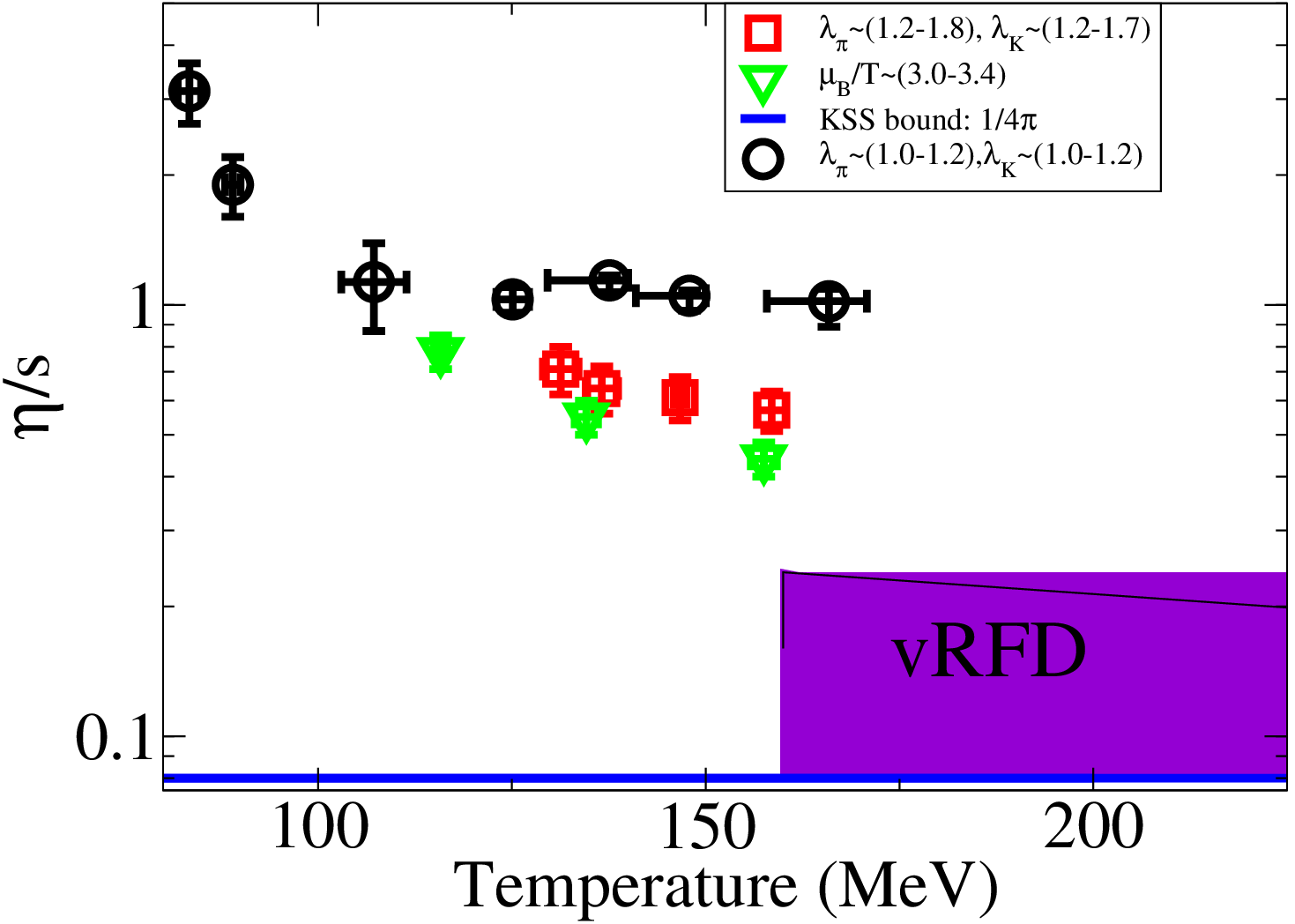}
\caption{LEFT:$\eta/s$ in full chemical and kinetic equilibrium.  Also illustrated are the results for 3 flavor pQCD and chiral pions.  RIGHT: $\eta/s$ at non-unit fugacities and finite baryo-chemical potential.}
\end{figure}

\section{Effect of Chemical Nonequilibrium on $\eta/s$}\label{}

However, calculating $\eta/s$ in full kinetic and chemical equilibrium,  as mentioned earlier, is not satisfactory: Although a statistical model analysis of particle yields and ratios at the RHIC indicate a chemical freeze-out temperature near $T_{\rm chem}\sim 160$ MeV, hydrodynamic calculations indicate that a kinetic freeze-out temperature of $T_{\rm kin}\sim 130$ MeV is required to describe the momentum distributions of final state hadrons \cite{freezeout}.  In order to mimic this situation in our simulations, we initialize our simulations with off-equilibrium yields and induce a non-unit pion/kaon fugacity.  The measurement of the viscosity is then made before the system relaxes back into chemical equilibrium, and the fugacity is extracted from particle ratios at the time of measurement.  Likewise a finite baryochemical potential can be induced by initializing the system with a surplus of baryons relative to antibaryons.  Our results for a system with nonunit fugacities (red points) and at finite baryochemical potential (green points) are juxtaposed with results for $\eta/s$ in full chemical and kinetic equilibrium (black points) in the right panel of Figure 2.  This tells us of a potentially very rich structure for $\eta/s$ in a heavy ion collision: as the deconfined phase with low $\eta/s$ cools and reaches $T_c$, $\eta/s$ could possibly rise sharply, then decrease again as the hadronic phase evolves out of chemical equilibrium before $\eta/s$ rises as it approaches kinetic freezeout \cite{ndprl}.  Work is now in progress for calculating the bulk viscosity in our model, since that is also an effect that is needed for viscous hydrodynamics calculations \cite{vreport,denicol}.



\section*{Acknowledgments} This work was supported by DOE grants DE-FG02-03ER41239 and DE-FG02-05ER41367.  We wish to thank Berndt M\"uller, Ulrich Heinz, Jorge Casalderrey-Solana, Derek Teaney, Giorgio Torrieri, and Pasi Huovinen for helpful discussions and suggestions.  ND thanks the organizers of the QM 2009 conference for local support.  

\end{document}